\documentclass[11pt,fleqn]{article}
\usepackage{latexsym}
\usepackage{amssymb}
\usepackage{amsmath}
\usepackage[hypertex]{hyperref}
\usepackage{graphicx}

\begin{document}

\begin{flushright}
hep-ph/0412200 \\
UT-04-24\\
\end{flushright}

\vspace{1.5cm}

\begin{center}
\mbox{\bf\LARGE A Viable Supersymmetric Model} \\
\vspace*{4mm}
\mbox{\bf\LARGE with UV Insensitive Anomaly Mediation} \\

\vspace*{1.5cm}
{\large Masahiro Ibe$^a$, Ryuichiro Kitano$^b$, and Hitoshi
  Murayama$^{b,c,d}$} \\
\vspace*{0.5cm}

\mbox{$^a$\textit{Department of Physics, University of Tokyo, Hongo, Japan}} \\

\mbox{$^b$\textit{School of Natural Sciences, Institute for Advanced Study, 
Princeton, NJ 08540}} \\

\mbox{$^c$\textit{Department of Physics,
University of California, Berkeley, California 94720}} \\

\mbox{$^d$\textit{Theoretical Physics Group, Lawrence Berkeley National
Laboratory}}
\mbox{\textit{Berkeley, California 94720}} \\

\vspace*{0.5cm}

\end{center}

\vspace*{1.0cm}

\begin{abstract}
We propose an electroweak model which is compatible with the UV
insensitive anomaly mediated supersymmetry breaking.
The model is an extension of the NMSSM by adding vector-like matter fields
which can drive the soft scalar masses of the singlet Higgs field
negative and the successful electroweak symmetry breaking is achieved.
Viable parameter regions are found to preserve perturbativity of all
the coupling constants up to the Planck scale.
With this success, the model becomes a perfect candidate of physics
beyond the standard model without the FCNC and CP problem. 
The cosmology is also quite interesting. The lightest neutralino is
the wino which is a perfect cold dark matter candidate assuming the
non-thermal production from the gravitino decay. There is no gravitino
problem because it decays before the BBN era, and thus the thermal
leptogenesis works. The cosmological domain wall problem inherent in
the NMSSM is absent since the $Z_3$ symmetry is broken by the QCD
instanton effect in the presence of the vector-like quarks.
We also briefly comment on a possible solution to the strong CP problem
 {\it \`{a} la} the Nelson-Barr mechanism.

\end{abstract} 

\newpage

\section{Introduction}

Supersymmetric (SUSY) standard model (SM) is an attractive framework
which realizes the standard model at low energy and suggests unification
of gauge interactions at high energy. The SUSY breaking triggers the
electroweak symmetry breaking while stabilizing the Higgs potential
against the radiative corrections.

Despite the great success in the gauge and Higgs sectors, the matter
sector is problematic.
With generic SUSY breaking terms in the Lagrangian, the prediction of
small CP violation and Flavor Changing Neutral Current (FCNC) in the SM
may be destroyed by the new interactions among the fermions and their
scalar partners.
The problems suggest the special features of the SUSY breaking, which
are flavor blind and CP conserving.

The gravitino, the SUSY partner of the graviton, also causes a problem
in cosmology. In the gravity mediated SUSY breaking scenario, the
gravitino mass is of the order of TeV and it decays during big-bang
nucleosynthesis (BBN), destroying the successful prediction of the
abundance of the light elements. In order to avoid the problem, the
reheating temperature of the universe has to be lower than $10^{6}$~GeV
\cite{Kawasaki:2004yh}, which is too low for many baryogenesis scenarios
with out-of-equilibrium decays of heavy particles.

An interesting possibility which solves all those problems automatically
is the anomaly mediated SUSY breaking scenario 
\cite{Randall:1998uk,Giudice:1998xp}. 
The anomaly mediation is the purely gravitational
mediation mechanism of the SUSY breaking and is realized once the direct
couplings between the hidden and visible sector fields are
suppressed. The SUSY breaking only appears with the conformal anomaly of
the theory and therefore it is insensitive to the ultraviolet (UV)
physics and respect all the accidental symmetries in low energy such as
CP and flavor conservation in the SM.
Moreover, the soft SUSY breaking terms are suppressed by the loop factor
$1/(4 \pi)^2$ compared to the gravitino mass because of its quantum
origin, and therefore the gravitino mass is naturally of
$O(100~{\rm TeV})$. Such a heavy gravitino decays before the BBN era,
opening a window of baryogenesis at high temperatures
\cite{Ibe:2004tg}.

The UV insensitivity enables us to calculate all the SUSY breaking terms
with known coupling constants. Unfortunately, the high predictability
immediately excludes the pure anomaly mediation since the scalar leptons
turn out to be tachyonic.
There have been several proposals to cure the problem. 
For example, to go off the trajectory of the anomaly mediation, one
considers non-decoupling effects by using flat directions 
\cite{Pomarol:1999ie, Katz:1999uw, Abe:2001cg, Okada:2002mv} or low
energy thresholds \cite{Nelson:2002sa}.
The introduction of SM non-singlet particles which feel SUSY breaking
directly from the hidden sector is shown to modify the spectrum through
two-loop diagrams \cite{Chacko:2001jt}.
Also, attempts to modify the trajectory of the anomaly mediation itself
by adding new interactions have been made 
\cite{Chacko:1999am, Allanach:2000gu, Anoka:2003kn, Shafi:2004cf}.
As modification of the initial conditions, several kinds of scenario
have been proposed such as inclusion of the K{\" a}hler anomaly
\cite{Bagger:1999rd}, adding $D$-terms
\cite{Jack:2000cd, Arkani-Hamed:2000xj, Anoka:2004vf}, and adding the
boundary interactions in the extra-dimensional setup
\cite{Kaplan:2000jz}.
The admixture of the gauge mediation and the anomaly mediation is
recently considered in the context of the conformal sequestering
scenario \cite{Sundrum:2004un}.
Among those various modifications, adding $D$-terms is clearly the safest
since it has no danger of reintroductions of the CP or FCNC problems,
and moreover it is shown to preserve the UV insensitivity which ensures
the flavor blind and CP conserving soft terms in low energy even in the
presence of the flavor changing or CP violating interactions in high
energy~\cite{Jack:2000cd, Arkani-Hamed:2000xj}.

Once we realize the UV insensitive anomaly mediation by the $D$-term
modification, the next step is to consider the electroweak symmetry
breaking to see if it is possible to have the desired vacuum
\cite{Anoka:2004vf, Kitano:2004zd}. When all the
SUSY breaking terms are calculable, it is highly non-trivial to have the
correct vacuum expectation values (VEV) of the Higgs fields. In
Ref.\cite{Kitano:2004zd}, it has been examined and found that the
minimal SUSY standard model (MSSM) does not have a stable vacuum unless
the $\tan \beta$ parameter, the ratio of two VEVs of the Higgs fields
($\langle H_2^0 \rangle / \langle H_1^0 \rangle$), is less than unity.
In this case, the top-Yukawa coupling constant is very large and blows
up right above the stop mass scale. Also, in the next to MSSM (NMSSM)
the prediction to the Higgsino mass is too small in all the region of
the parameter space. It is caused by the fact that the singlet Higgs
field $S$ only has Yukawa interactions which are asymptotically
non-free.  The soft mass squared is likely to be positive in that case,
resulting in the small VEV of $S$.
A successful model is found with linear term of the singlet field in
the superpotential, which happens to be the low energy effective
theory of the minimal SUSY fat Higgs model \cite{Harnik:2003rs}.

In this paper, we reexamine the NMSSM with $D$-term modified anomaly
mediation by introducing additional vector-like
quarks~\cite{Dine:1993yw, Agashe:1997kn, deGouvea:1997cx,
Chacko:1999am}. We find that the coupling between the singlet Higgs
field $S$ and the vector-like quarks can modify the anomalous dimension
of $S$ significantly, and make the soft mass squared of $S$ small enough
to acquire a large VEV and thus there is no Higgsino mass problem.

This paper is organized as follows: 
In the next section, we review the UV insensitive anomaly mediation. We
discuss the problems in the MSSM and NMSSM with the UV insensitive
anomaly mediation in Section \ref{sec:ewsb}. We extend the NMSSM with
additional vector-like matter fields and examine the electroweak
symmetry breaking in Section \ref{sec:vector}, and discuss
phenomenological and cosmological issues. Section \ref{sec:conclusion}
is devoted to conclusions.

\section{UV insensitive anomaly mediation}

Anomaly mediation is realized once we obtain a sequestered K{\" a}hler
potential in the supergravity action.
Since the fields in the visible sector can feel the SUSY breaking only
through the gravitational interaction, the only source of the soft terms
is the $F$-component of the gravity multiplet.
In the superconformal formulation of the supergravity
Lagrangian~\cite{Cremmer:1978hn, Cremmer:1982en, Kugo:cu}, the
$F$-component is the auxiliary component of the chiral compensator
multiplet $\Phi$ with which Lagrangian possesses superconformal
symmetry.
The supergravity Lagrangian is obtained by fixing the value of the
components of $\Phi$ which breaks the superconformal symmetry explicitly
down to the super-Poincar{\' e} symmetry.
It is clear in this construction that the SUSY breaking appears only
with the violation of the conformal symmetry.
At the classical level, therefore, the soft scalar masses, the scalar
cubic couplings ($A$-terms), and the gaugino masses vanish, and all
those terms appear at quantum level with conformal anomaly. Explicitly,
with the anomalous dimension $\gamma_i$ of the chiral superfield $Q_i$
and the beta function $\beta_A$ of the gauge interaction labeled by $A$,
soft terms at the scale $\mu$ are given by
\begin{eqnarray}
        A_{ijk} = - \lambda_{ijk} ( \gamma_i + \gamma_j + \gamma_k )
        m_{3/2}\ ,\ \ \ \tilde{m}_i^2 = \frac{1}{2} \frac{d \gamma_i}{d
        \ln \mu} m_{3/2}^2\ ,\ \ \ m_{\lambda}^A = \frac{\beta_A}{g_A}
        m_{3/2}\ , \label{eq:soft}
\end{eqnarray}
where $\lambda_{ijk}$ and $g_A$ are the Yukawa and gauge coupling
constants, respectively. The mass parameter $m_{3/2}$ is the gravitino
mass which represents the $F$-component of the compensator multiplet.
The above soft terms are defined by
\begin{eqnarray}
        {\cal L}_{\rm soft} = - (A_{ijk} q_i q_j q_k + h.c.)  -
        \tilde{m}_i^2 |q_i|^2 - \frac{1}{2} m_\lambda^A \bar{\lambda}
        \lambda \ ,
\end{eqnarray}
with $q_i$ and $\lambda$ being the scalar component of $Q_i$ and
gauginos, respectively.
Of interest is that the soft terms at the scale $\mu$ are described by
$\gamma$ and $\beta$ at that scale and hence do not depend on the high
energy physics. The UV insensitivity is a phenomenologically desirable
feature since it solves the SUSY FCNC and CP problem automatically.

It had been thought that sequestered K\"ahler potential is difficult to
achieve in realistic model of quantum gravity. For instance, string
theory tends to give rise to many moduli fields who can mediate
additional supersymmetry breaking effects at the tree-level or one-loop
level, which dominate over the anomaly-mediated
contributions~\cite{Anisimov:2001zz, Anisimov:2002az}.  In
Ref.~\cite{Giudice:1998xp}, the absence of such fields was explicitly
assumed.\footnote{The paper discussed another possibility that the
gaugino mass is generated at one-loop level by the anomaly-mediation
effect whereas the scalar masses are not suppressed, realizing the split
SUSY scenario~\cite{Arkani-Hamed:2004fb,Wells:2004di}.}
The physical separation of hidden and observable sectors along
an extra dimension was used in Ref.~\cite{Randall:1998uk} to justify the
sequestered form of the K\"ahler potential.  This, however, is not
immune to the problem because there may be light bulk scalars such as
radion.  Only recently, a concrete mechanism to fix all of the moduli
fields was proposed~\cite{Kachru:2003aw}.  Moreover, the physical
separation was shown not necessary to achieve sequestered K\"ahler
potential if the hidden sector is nearly conformal~\cite{Luty:2001jh,
Luty:2001zv}, even though the hidden sector needs to be of a special
type~\cite{Dine:2004dv}.  Therefore achieving sequestered form of the
K\"ahler potential does not appear to be an insurmountable problem any
more.

An obvious problem of the framework is the tachyonic sleptons. The
contribution from the gauge interaction to the scalar masses
$\tilde{m}_i^2$ is positive (negative) for asymptotically free
(non-free) gauge interaction. Since the sleptons only have SU(2)$_L$ 
and U(1)$_Y$ gauge interactions which are both asymptotically non-free,
the scalar masses squared of the sleptons turn out to be negative with
neglecting Yukawa coupling constants of the leptons.

However, we can easily solve the problem when we gauge the $B-L$
symmetry in the MSSM \cite{Arkani-Hamed:2000xj}. 
For example, if a U(1)$_A$ gauge symmetry in the hidden sector acquire
the $D$-term, the kinetic mixing term between U(1)$_A$ and U(1)$_{B-L}$
induces the $D$-term of U(1)$_{B-L}$.
%
%
%
The $D$-term of U(1)$_Y$ may also be generated in the same way.
Remarkably those two $D$-terms can provide sufficiently large positive
contributions for the scalar masses squared of both the left- and
right-handed sleptons. Moreover, the most significant feature is that
the modification with $D$-terms preserves the UV insensitivity of the
soft terms \cite{Arkani-Hamed:2000xj}.

The introduction of $D$-terms modifies the soft terms in the following
way.
\begin{eqnarray}
       \tilde{m}_i^2 = \frac{1}{2} \frac{d \gamma_i}{d \ln \mu} m_{3/2}^2
       - Q_Y^i D_Y - Q_{B-L}^i D_{B-L}\ ,
       \label{eq:soft+D}
\end{eqnarray}
where $Q_Y^i$ and $Q_{B-L}^i$ are the hypercharge and the $B-L$ charge
of the corresponding superfield. The gaugino masses and the $A$-terms
are not modified.
Neglecting the Yukawa coupling constants, we obtain the slepton masses
as follows:
\begin{eqnarray}
 m_{\tilde{l}}^2 = 
\left(
-\frac{11}{2} g_Y^4 -\frac{3}{2} g_2^4
\right) M^2
+ \frac{1}{2} D_Y
+ D_{B-L}\ ,
\end{eqnarray}
\begin{eqnarray}
 m_{\tilde{e}^c}^2 = 
-22 g_Y^4 M^2
- D_Y
- D_{B-L}\ ,
\end{eqnarray}
where $g_Y$ and $g_2$ are the U(1)$_Y$ and SU(2)$_L$ gauge coupling
constants, respectively, and $M=m_{3/2}/(4\pi)^2$. The additional
$D$-term contributions are positive for both of the sleptons when $D_Y <
- D_{B-L} < D_Y / 2 < 0$.

The $D$-terms can be obtained in a consistent way with grand unified
theories (GUT), even though U(1)$_Y$ $D$-term becomes gauge non-singlet.
For example, in SO(10) grand unified theories, U(1)$_Y$ and U(1)$_{B-L}$
are both subgroups of SO(10). The kinetic mixing terms between the
U(1)$_A$ gauge field and those U(1)'s in this case are generated after
the GUT breaking assuming the presence of the following term:
\begin{eqnarray}
 {\cal L} \ni \int d^2 \theta 
\frac{ \Sigma^K }{M_{\rm Pl} } ( W_{U(1)_A} )^\alpha 
( W_{\rm SO(10)}^K )_\alpha 
+ h.c. \ .
\end{eqnarray}
Here, $\Sigma^K$ is a chiral superfield of 45 dimensional
representation, which breaks SO(10) into SU(3)$_C$ $\times$ SU(2)$_L$
$\times$ U(1)$_Y$ $\times$ U(1)$_{B-L}$.
Provided the $D$-term of U(1)$_A$ is generated at one-loop level, the
above mixing induces the $D$-terms of U(1)$_Y$ and U(1)$_{B-L}$ of the
order of $(4\pi)^2 (M_{\rm GUT} / M_{\rm Pl}) M^2 $ which is numerically
similar size to the sfermion soft masses squared of $O(M^2)$.
Alternatively, in the models with orbifold unification
\cite{Kawamura:2000ev, Altarelli:2001qj, Hall:2001pg}, we can simply
write down a kinetic mixing term between U(1)$_Y$ and U(1)$_{B-L}$ gauge
fields at the boundary.

\section{Electroweak symmetry breaking in the MSSM and NMSSM}       
\label{sec:ewsb}

We now proceed to consider the electroweak symmetry breaking with the
$D$-term modified anomaly mediation.
Whether we obtain correct vacuum is quite non-trivial since all the soft
terms are calculable with known coupling constants.

\subsection{MSSM}
In Ref.\cite{Kitano:2004zd} it has been shown that the correct vacuum
(correct $Z$ boson mass) is realized only when $\tan \beta \lesssim 0.3$
which causes the Landau pole of the top-Yukawa coupling constant just
above the SUSY breaking scale.
The problem is caused by the explicit violation of the conformal
symmetry by the $\mu$-term, i.e., $W \ni \mu H_1 H_2$. Because it is a
tree-level violation, the $B$-term associated with the $\mu$-term is not
suppressed by the loop factor and hence is large by a factor of $(4
\pi)^2$ compared to other soft terms. Such a huge $B$-term requires
large Yukawa couplings to fine-tune the VEV of the Higgs fields by
enhancing soft masses of $H_1$ and $H_2$ through Eq.(\ref{eq:soft}).

\begin{figure}[t]
\begin{center}
\includegraphics[height=6.5cm]{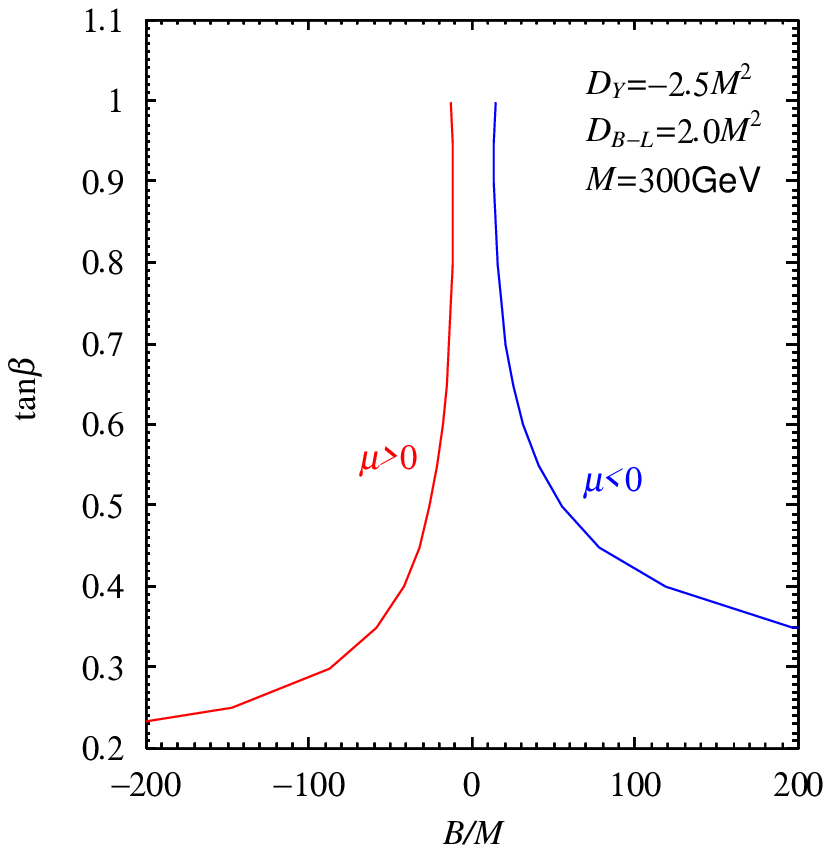}
\includegraphics[height=6.5cm]{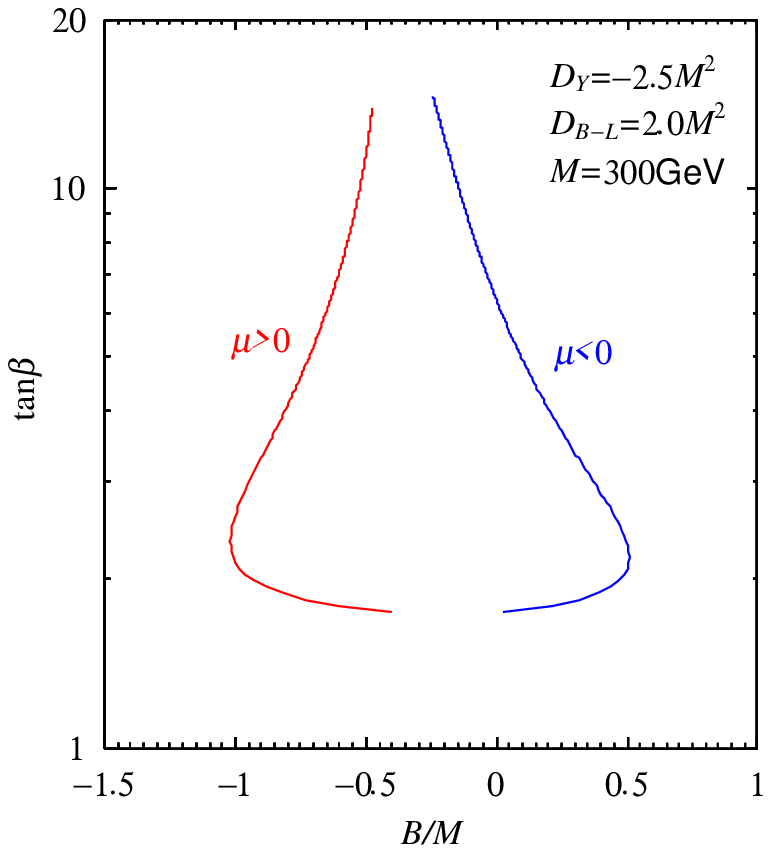} 
\end{center}
\caption{The
relation between $B$-parameter and $\tan \beta$ is shown. We
fixed the gravitino mass parameter ($M=m_{3/2}/(4 \pi)^2$) to be
300GeV and $D$-terms ($D_Y = - 2.5 M^2$ and $D_{B-L} = 2.0 M^2$).} 
\label{fig:B-tanbeta}
\end{figure}

There is a way of suppressing the $B$-term by using an accidental
cancellation among different sources of the $\mu$- and $B$-terms. For
example, if we have the following terms in the K{\" a}hler and the
superpotential,
\begin{eqnarray}
        K \ni c H_1 H_2 + h.c.\ ,\ \ \ W \ni \tilde{\mu} H_1 H_2 \ ,
         \label{eq:kahler}
\end{eqnarray}
the $\mu$- and $B$-terms from the SUSY breaking are given by
\begin{eqnarray}
        \mu = c m_{3/2} + \tilde{\mu}\ ,\ \ \ 
         B \mu = m_{3/2} ( - c m_{3/2} + \tilde{\mu} )\ .
         \label{eq:mu_and_B}
\end{eqnarray}
Since the relative sign between the two contributions is different for
the $\mu$- and $B$-terms, there is a possibility to have small $B$-term
by carefully tuning the $c$ and $\tilde{\mu}$ parameters.
In this situation, we can think of the $B$-parameter as a free
parameter. We plot in Fig.\ref{fig:B-tanbeta} the $B$-parameter
dependence of $\tan \beta$ with fixing $M$, $D_Y$, and $D_{B-L}$.
We have included the one-loop correction to the Higgs potential from the
(s)top, (s)bottom, and (s)tau loop diagrams, and imposed the stability
of all the scalar particles.
For large $|B|/M$, such as $(4 \pi)^2 \sim 158$, we need a small value
of $\tan \beta$ so that the top-Yukawa coupling constant gives large
contribution to the Higgs potential. The solutions with $\tan \beta
\gtrsim 1$ are found with small values of $B$. Therefore, even in the
MSSM, we could achieve the successful electroweak symmetry breaking with
maintaining perturbativity up to the Planck/GUT scale at the expense of
a fine-tuning of order $10^{-2}$ in Eq.(\ref{eq:mu_and_B}).

Unfortunately, the model has a potential danger to introduce a new CP
phase since $c$ and $\tilde{\mu}$ are independent complex
parameters. Although the reintroduction of the SUSY CP problem upsets
the motivation of the UV insensitive anomaly mediation, there is an
interesting observation. As we discuss later in Sec.~\ref{sec:strongCP}, the
anomaly mediation is a good framework for solving the strong CP problem
by the scenario of the spontaneous CP violation
\cite{Nelson:1983zb,Barr:qx}.
Since the scenario assumes the exact CP invariance in the Lagrangian,
$c$ and $\tilde{\mu}$ are real parameters if the fields which break the
CP invariance do not couple with the Higgs fields.
In this sense, we claim that the MSSM is a viable model with the UV
insensitive anomaly mediation once we obtain small $B$-parameters.

\subsection{NMSSM}
Apart from the benefit of providing a solution to the $\mu$-problem, the
extension to the NMSSM is well-motivated in the context since the
conformal symmetry is not violated at tree level, and hence there is no
complication caused by the $(4 \pi)^2$ enhanced soft parameters.
The superpotential of the Higgs sector in the NMSSM is given by
\begin{eqnarray}
W = \lambda S H_1 H_2 + \frac{h}{3} S^3\ ,
\end{eqnarray}
where $S$ is the gauge singlet Higgs field whose VEV plays a role of the
$\mu$-parameter in the MSSM.

The electroweak symmetry breaking in the NMSSM is, however, not
successful since the effective $\mu$-parameter of $\lambda \langle S
\rangle$ is too small (at most a few GeV), resulting in unacceptably light
Higgsinos~\cite{Kitano:2004zd}. The problem is caused by the positivity
of the soft mass squared for the singlet Higgs $m^2_S$.
The $m^2_S$ parameter is given by the formula:
\begin{eqnarray}
        m^2_S = \frac{1}{2}
         (4 \pi)^2 
         \frac{d}{d \ln \mu} 
         \left( 
          2 \lambda^2 + 2 h^2
              \right) M^2\ .
         \label{eq:mSsq}
\end{eqnarray}
Ignoring the SU(2)$_L$ and U(1)$_Y$ gauge interactions, $\lambda$
and $h$ are asymptotically non-free, and thus $m_S^2$ is positive.
The positivity indicates the stability of the origin of the potential,
and therefore the VEV of $S$ is only induced by the small shift of the
origin through the linear term $\lambda A_\lambda \langle H_1^0 \rangle \langle
H_2^0 \rangle S$ in the potential.
Although the linear term can be enhanced by increasing the coupling
$\lambda$ and/or $h$, it is of no help since that also enhances $m_S^2$.

\section{Modified NMSSM with Vector-like Matter Fields}
\label{sec:vector}

\subsection{The Model}
Having understood the problems in the MSSM and the NMSSM, we now
consider a model with vector-like matter fields. The problem in the NMSSM is
caused by the high predictability of the anomaly mediation. The soft
terms are calculable once we fix a model. Particularly the large
positive $m_S^2$ parameter is problematic if we assume the minimal
interactions. However, the interactions of $S$ are phenomenologically
unknown and we can easily modify the soft terms of the
singlet. As an example, we consider a model with vector-like matter
fields which couple to the $S$ field.

A similar model has been considered in Ref.\cite{Chacko:1999am} in the
context of the anomaly mediation without $D$-terms.
It was pointed out that the calculability of the soft terms still makes
the electroweak symmetry breaking difficult even in the presence of $S$
and extra vector-like fields by the following reason.
The $\tan \beta$ parameter is determined by the difference between two
soft masses of the Higgs doublets, $m_{H_1}^2$ and $m_{H_2}^2$, which
only depends on the top Yukawa coupling constant $f_t$ in that
model. Therefore, the top-quark mass $m_t ( = f_t v \sin \beta ) $ is
predicted rather than an input parameter.
Unfortunately, successful electroweak symmetry breaking turns out to
require unacceptably small $m_t$ such as less than 145~GeV.
They discussed a further extension of the model to overcome the situation
by introducing three extra singlet superfields and appropriate
interaction terms.
In contrast, we show that the correct vacua are easily found in the
model with $D$-terms. Since $D_Y$ contributes to $m_{H_1}^2$ and
$m_{H_2}^2$ with opposite signs, we do not have the unwanted correlation
between $\tan \beta$ and $f_t$ anymore.
We can find phenomenologically viable parameter regions without
introducing extra singlets except for $S$.

We introduce a pair of new chiral superfields $D$ and $\bar{D}$ which
have quantum numbers of $({\bf 3,1)_{-1/3}}$ and $({\bf
  \bar{3},1)_{1/3}}$ under the SM gauge group, and also $L$ and
$\bar{L}$ of $({\bf 1,2)_{-1/2}}$ and $({\bf 1,2)_{1/2}}$ so that the
extra matter fields form complete SU(5) representations ${\bf
  5}+\overline{\bf 5}$ to ensure the gauge coupling unification.
We can write down the following superpotential:
\begin{eqnarray}
 W = \lambda S H_1 H_2 + \frac{h}{3} S^3 
+ k_D S \bar{D} D + k_L S \bar{L} L \ .
\label{eq:superp}
\end{eqnarray}
The introduction of the interactions between $S$ and vector-like matter fields
does not cause a new CP violation, since the coupling constants $\lambda$,
$h$, $k_D$, and $k_L$ can be made real without loss of generality by
appropriate field redefinitions.
We do not assume the presence of the direct couplings of $D$ and $L$
with the ordinary quarks and leptons, which may be forbidden by a U(1)
symmetry or its discrete subgroup under which only the vector-like
matter fields transform.
The additional interactions in Eq.(\ref{eq:superp}) modify the $m_S^2$
parameter from Eq.(\ref{eq:mSsq}) to (see Appendix for the RGEs)
\begin{eqnarray}
        m^2_S = 
         \frac{1}{2}
         (4 \pi)^2 
         \frac{d}{d \ln \mu} 
         \left( 
          2 \lambda^2 + 2 h^2
          + 3 k_D^2 + 2 k_L^2
              \right) M^2\ .
\label{eq:mSnew}
\end{eqnarray}
Since $k_D$ is asymptotically free in a wide range of parameter space
due to the strong interaction of $D$ and $\bar{D}$, the negative
contribution may make $m^2_S$ negative and large in its absolute value
such that we obtain large values of $\langle S \rangle$.

\subsection{Electroweak symmetry breaking}
Let us evaluate the VEV of the Higgs fields by minimizing the potential
with soft terms calculated by Eqs.(\ref{eq:soft}) and (\ref{eq:soft+D}).
The potential for the neutral components of the Higgs fields are given
by
\begin{eqnarray}
 V &=& ( m_{H_1}^2 + | \lambda S |^2 ) |H_1^0|^2
   + ( m_{H_2}^2 + | \lambda S |^2 ) |H_2^0|^2
   + m_S^2 |S|^2 \nonumber \\
   &&
   + |\lambda H_1^0 H_2^0 + h S^2|^2 
   + ( A_\lambda S H_1^0 H_2^0 + \frac{A_h}{3} S^3 + h.c.) \nonumber \\
   &&
   + \frac{1}{8} (g_Y^2 + g_2^2) (|H_1^0|^2 - |H_2^0|^2)^2\ .
\end{eqnarray}
The soft terms are given as follows:
\begin{eqnarray}
 m_{H_1}^2 = \frac{1}{2} (4 \pi)^2 \frac{d}{d \ln \mu}
\left(
f_\tau^2 + 3 f_b^2 + \lambda^2 - \frac{1}{2} g_Y^2 - \frac{3}{2} g_2^2 
\right) M^2 
+ \frac{1}{2} D_Y \ ,
\end{eqnarray}
\begin{eqnarray}
 m_{H_2}^2 = \frac{1}{2} (4 \pi)^2 \frac{d}{d \ln \mu}
\left(
3 f_t^2 + \lambda^2 - \frac{1}{2} g_Y^2 - \frac{3}{2} g_2^2 
\right) M^2
- \frac{1}{2} D_Y \ ,
\end{eqnarray}
\begin{eqnarray}
 A_\lambda = - \lambda 
\left(
3 f_t^2 + f_\tau^2 + 3 f_b^2 
+ 4 \lambda^2 + 2 h^2
+ 3 k_D^2 + 2 k_L^2
- g_Y^2 - 3 g_2^2 
\right) M\ ,
\end{eqnarray}
\begin{eqnarray}
 A_h = - h
\left(
6 \lambda^2 + 6 h^2
+ 9 k_D^2 + 6 k_L^2
\right) M\ ,
\end{eqnarray}
and $m_S^2$ is given in Eq.(\ref{eq:mSnew}). Here, $f_t$, $f_b$, and
$f_\tau$ are the Yukawa coupling constants of the top quark, the bottom
quark, and the tau lepton, respectively, and we ignored those for the
first and the second generations.
The scale dependence of the Yukawa and gauge coupling constants are
given in Appendix \ref{app:rge}.
Notice that $k_D$ enhances the $A$-terms while suppressing $m_S^2$ in
Eq.(\ref{eq:mSnew}), which makes it possible for $S$ to acquire a large
VEV.
The minimization conditions with respect to the three Higgs fields $S$,
$H_1^0$, and $H_2^0$ are given by
\begin{eqnarray}
 A_\lambda S + \lambda h S^2 
&=& - \frac{\sin 2 \beta}{2} (m_1^2 + m_2^2 + \lambda^2 v^2) 
\label{eq:const-1}\\
 m_Z^2
&=& - \frac{m_1^2 - m_2^2}{\cos 2 \beta} - (m_1^2 + m_2^2)\\
 0
&=& \lambda^2 v^2 + 2 h^2 S^2 + ( \lambda h + \frac{A_\lambda}{2S}) v^2 \sin 2 \beta
+ A_h S + m_S^2 \ ,
\label{eq:const-3}
\end{eqnarray}
where $m_1 \equiv m_{H_1}^2 + \lambda^2 S^2$ and $m_2 \equiv m_{H_2}^2 + \lambda^2 S^2$, and $v^2 =\langle H_1^0 \rangle^2 + \langle H_2^0 \rangle^2 $.


%
We numerically solve the Eqs.(\ref{eq:const-1}--\ref{eq:const-3}) while
fixing $v = 174$~GeV.
Once we postulate values of $k_D$, $k_L$, $h$, $\tan \beta$, and $M$,
the three equations determine the sizes of $\lambda$, $S$, and $D_Y$.
\begin{figure}[t]
\begin{center}
 \includegraphics[height=6.5cm]{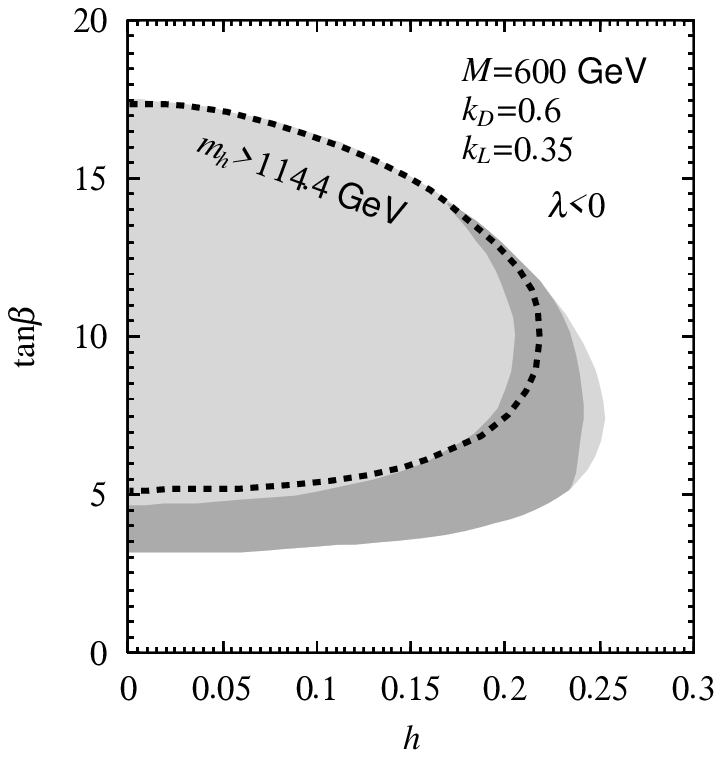}\hspace*{5mm}
 \includegraphics[height=6.5cm]{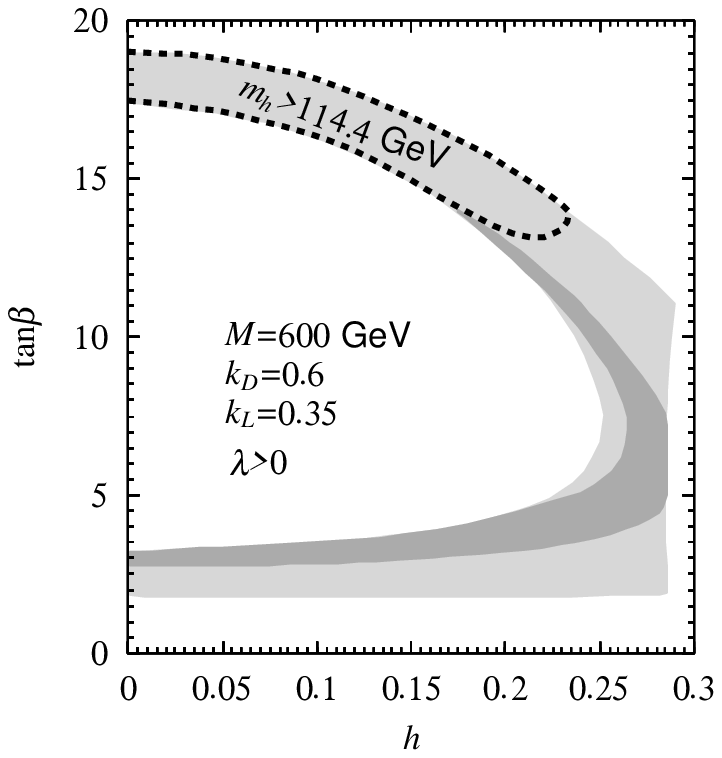} \caption{The
 parameter regions where stable solutions are found in the modified NMSSM
 are plotted (the light gray shaded region). Left and right figures
 correspond to the solutions with $\lambda<0$ and $\lambda > 0$,
 respectively.
The LSP is a neutralino in the dark shaded region and the lightest Higgs
boson is heavier than $114.4$~GeV inside the dashed line. The other
parameters are fixed to be $k_D=0.6$, $k_L=0.35$, and $M=600$~GeV.} 
\label{fig:solutions}
\end{center}
\end{figure}       
\begin{figure}
\begin{center}
 \includegraphics[height=6cm]{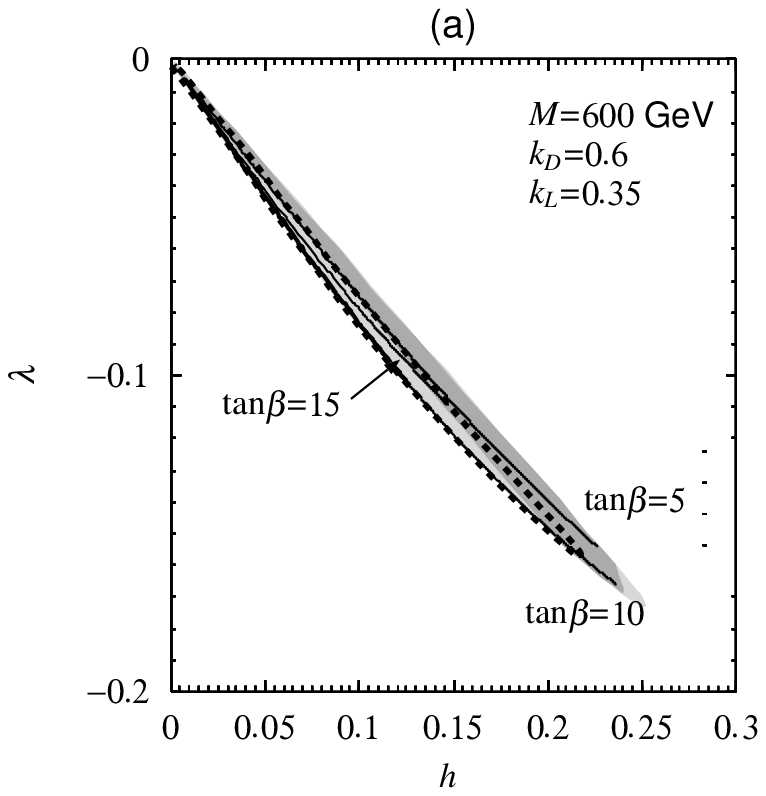}\hspace{5mm}
 \includegraphics[height=6cm]{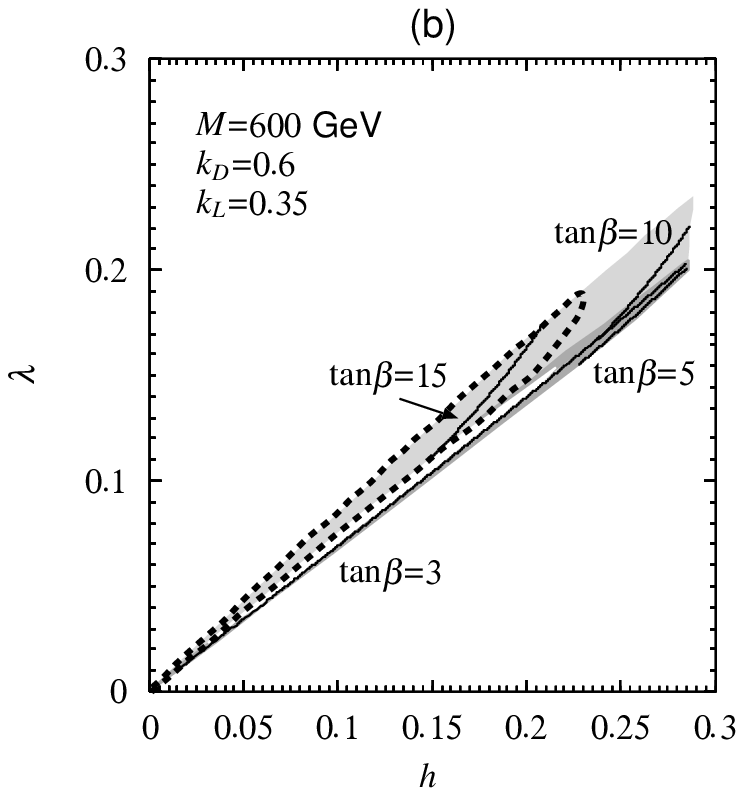}

 \hspace{-3mm}
 \includegraphics[height=6cm]{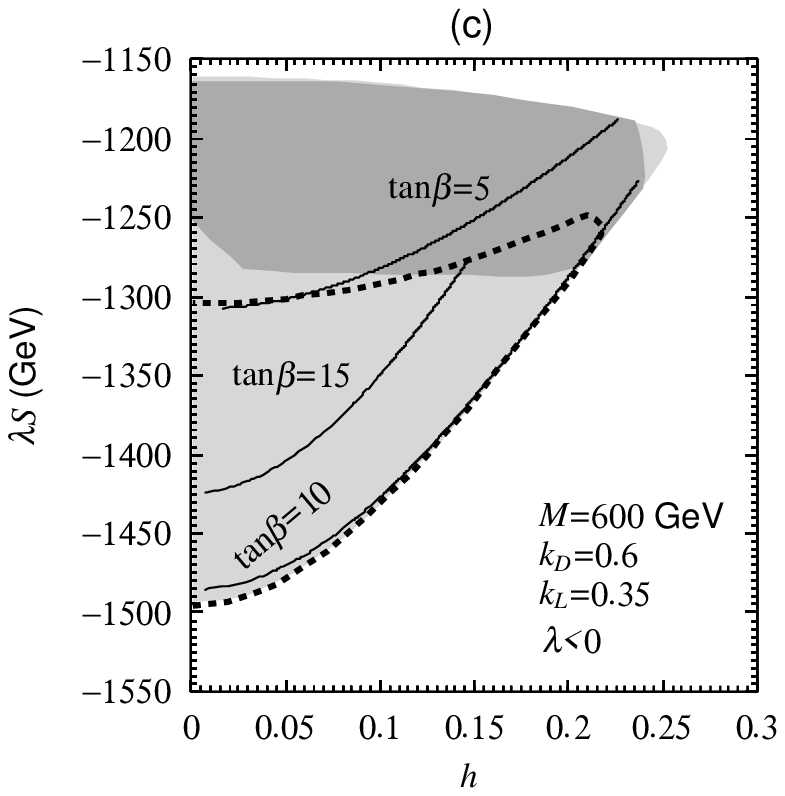}\hspace{3.5mm}
 \includegraphics[height=6cm]{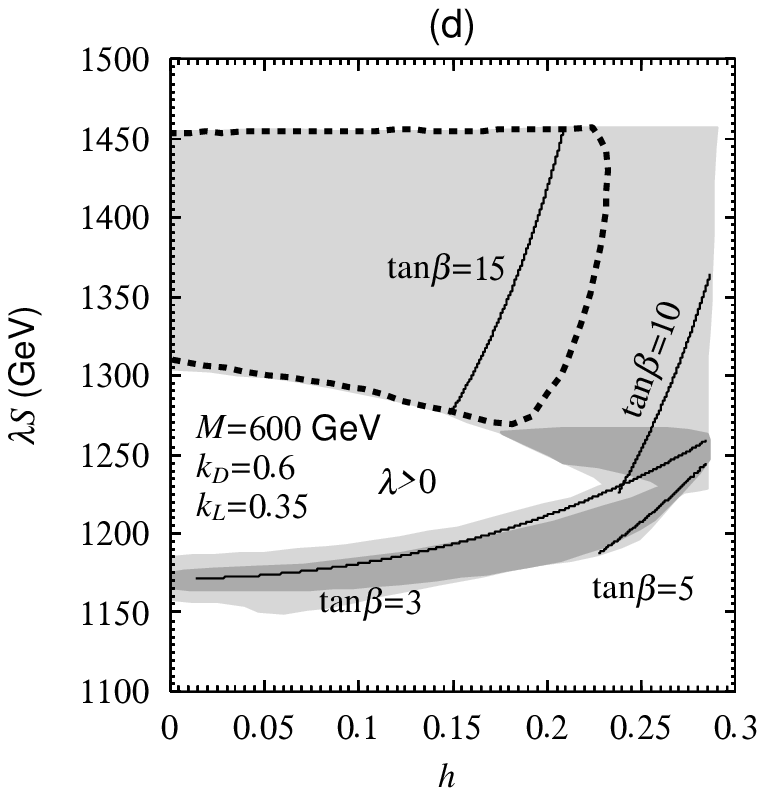}

 \hspace{1.5mm}
 \includegraphics[height=6cm]{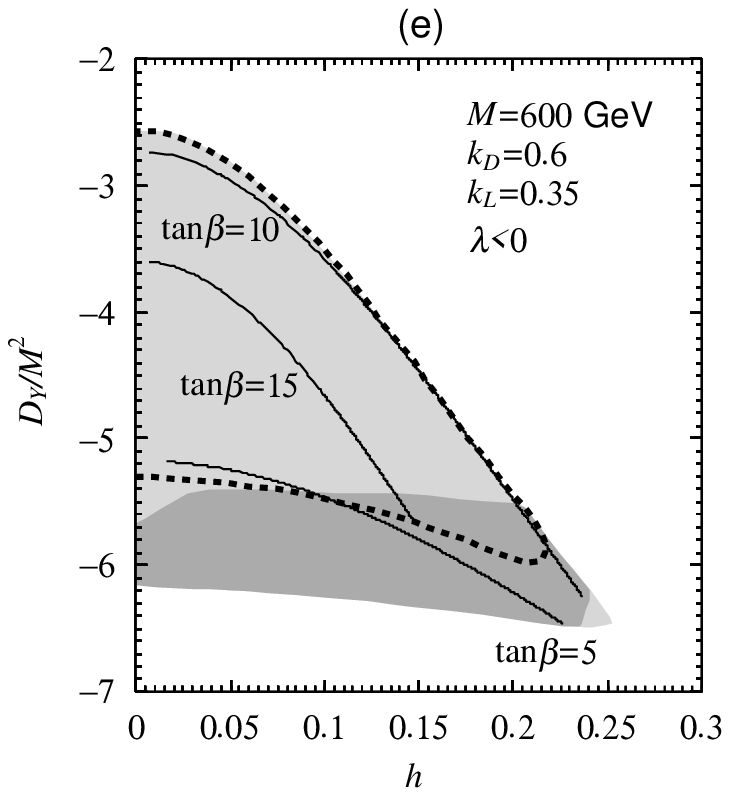}\hspace{7mm}
 \includegraphics[height=6cm]{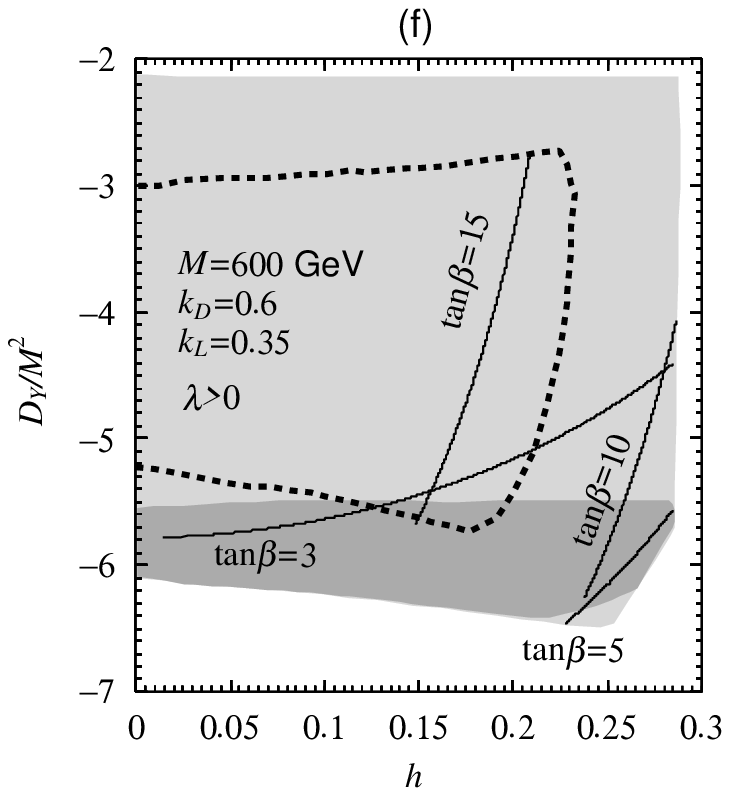} \caption{
   The regions with stable solutions are shown. Lines show the $h$
   dependence of $\lambda$ (a,b), $\lambda S$ (c,d), and $D_Y$ (e,f)
   with fixed values of $\tan \beta$. The neutralino LSP is realized
   in the dark shaded region. Left (a,c,e) and right (b,d,f) figures
   correspond to the solutions with $\lambda < 0$ and $\lambda >0$,
   respectively. The other parameters are fixed to be $M=600$~GeV,
   $k_D=0.6$, and $k_L=0.35$.  The overlaps of the Higgs mass
   constraint and the dark shaded region in (f) is fictitious as the
   parameter space is covered multiple times.  }
\label{fig:nmssm}
\end{center}
\end{figure}       
Stable minima are found with reasonable parameter sets as shown in the
gray shaded region in Fig.\ref{fig:solutions}. The left and right
figures correspond to solutions with negative and positive values of
$\lambda$, respectively.
The solutions are found for the range $1.5 \lesssim \tan \beta
\lesssim 19$ and $0 < h \lesssim 0.3$.
The parameters are fixed in the plots to be $k_D = 0.6$ and $k_L =0.35$
motivated by the unification at the GUT scale, and $M=600$~GeV. 
In the dark gray shaded region, the lightest SUSY particle (LSP) is a
neutralino (mostly the wino). The stop and/or stau are lighter than the
neutralino outside the region.
The experimental bound on the lightest Higgs boson mass \cite{:2003ih}
is satisfied inside the dashed line.
We can see the overlaps of the two regions for the
$\lambda < 0$ case.
We plot those solutions $\lambda$, $\lambda S$, and $D_Y / M^2$ as
functions of $h$ in Fig.\ref{fig:nmssm}.
Solutions are found in the gray shaded region and again the LSP is a
neutralino in the dark gray shaded region.  Lines correspond to various
values of $\tan \beta$.
We see in Fig.{\ref{fig:nmssm} (a) and (b)} that the values of
$|\lambda|$ and $h$ are restricted to be small such as less than
0.3. The large values are not preferred since those enhance $m_S^2$.
Fig.\ref{fig:nmssm}(c) and (d) show that we obtain large effective
$\mu$-parameter of the order of 1000~GeV such that the Higgsino is heavy
enough.
The negative values of $D_Y$ in Fig.\ref{fig:nmssm}(e) and (f) ensure
the positive contributions to the slepton masses squared by taking
$D_{B-L}$ in an appropriate range.

\subsection{Spectrum and phenomenology}
We show a mass spectrum at a sample point in the parameter space and
discuss phenomenological implications.
As a point with the neutralino LSP, we choose parameters such as
$h=0.21$ and $\tan \beta = 10$ in Fig.\ref{fig:solutions}.
The mass spectra are listed in Table \ref{tab:nmssm}. We take the $B-L$
charge of the extra-fields to be $D:1/3$ and $L:-1$, although those are,
in principle, arbitrary.

\begin{table}
\begin{center}
\begin{tabular}{|l|l||l|l|} 
\hline\hline
 ${h_1^0}$        & 115       &  ${\tilde{l}_L}$  & 630   \\
 ${h_2^0}$        & 464      & ${\tilde{e}_R}$  & 509   \\
 ${h_3^0}$        & 2743     & ${\tilde{\tau}_1}$  & 495\\
 ${A_1^0}$        & 462      & ${\tilde{\tau}_2}$  & 628\\
 ${A_2^0}$        & 3703      & ${\tilde{\nu}}$  & 625   \\
 ${H^\pm}$        & 470      & ${\tilde{\nu}_\tau}$  & 621   \\
 ${\chi_1^\pm}$   & 489     & ${\tilde{u}_L}$  & 1564   \\
 ${\chi_2^\pm}$   & 1277      & ${\tilde{u}_R}$  & 1397   \\
 ${\chi_1^0}$     & 489    & ${\tilde{t}_1}$  & 552   \\
 ${\chi_2^0}$     & 979      & ${\tilde{t}_2}$  & 1280   \\
 ${\chi_3^0}$     & 1273      & ${\tilde{d}_L}$  & 1566   \\
 ${\chi_4^0}$     & 1280      & ${\tilde{d}_R}$  & 2013   \\
 ${\chi_5^0}$     & 3478      & ${\tilde{b}_1}$  & 1256   \\
 ${\tilde{g}}$    & 1459      & ${\tilde{b}_2}$  & 1984 \\
\hline\hline
\end{tabular}
\hspace{1cm}
\begin{tabular}{|l|l|} 
\hline\hline
 $D$  & 4969 \\
 ${\tilde{D}_1}$  & 2294 \\
 ${\tilde{D}_2}$  & 6973 \\
 $L$  & 2898 \\
 ${\tilde{L}_1}$  & 1927 \\
 ${\tilde{L}_2}$  & 3560 \\
\hline\hline
\end{tabular}

\end{center}
\caption{A sample mass spectrum is shown in the unit of
GeV. We take $M=600$~GeV, $\lambda = -0.15$, $h = 0.21$, $D_Y =
-5.7M^2$, $D_{B-L} = 4.6M^2$, and $\tan \beta = 10$.  The $k_D$ and
$k_L$ coupling constants are taken to be $0.6$ and $0.35$, respectively.
The Higgs potential has a minimum with the correct size of the Higgs VEV
and $S=8.3$~TeV. We have included the one-loop correction to the
lightest Higgs boson mass.} \label{tab:nmssm}
\end{table}


The lightest Higgs boson $h_1^0$ is likely to be mainly composed by the
doublet Higgs fields $H_1^0$ and $H_2^0$. There is no significant mixing
between the singlet and the doublet Higgs boson because of the small
$\lambda$ parameter. Therefore properties of the lightest Higgs boson are
similar to those of the MSSM. In particular, $h_1^0$ is lighter than the
$Z$-boson at tree level, and hence the radiative corrections from the
(s)top loop diagrams are important to satisfy the experimental bound
\cite{Okada:1990vk,Ellis:1990nz,Haber:1990aw}.
The large stop masses are required in order to obtain the large
radiative corrections. As in the MSSM, this requires a relatively high
SUSY scale such as $M \gtrsim 600$~GeV.  
The charged Higgs boson mass for this parameter is well above the
limit from the $b\rightarrow s\gamma$, $m_{H^\pm} \gtrsim 350$~GeV
\cite{Gambino:2001ew}.\footnote{Because the solutions require $\lambda
  < 0$ and hence the chargino diagram constructively interferes with
  the charged Higgs boson diagram, the constraint is probably somewhat
  stronger than this in our case, but it is clear that a slightly
  higher $M$ can satisfy the limit if needed.  Detailed quantitative
  discussions on this constraint is beyond the scope of this paper.}

The LSP is mostly the wino, the SU(2)$_L$ gaugino.
The large Higgsino mass parameter $\lambda S$ indicates the small
gaugino-Higgsino mixing and thus the charged and neutral winos are
highly degenerate.
The dominant contribution to the mass splitting is the one-loop radiative
correction from diagrams with the gauge boson loops, and is estimated to
be 165~MeV for large $M$ \cite{Cheng:1998hc, Gherghetta:1999sw}.
In this case, the charged wino mainly decays into a charged pion and a
neutral wino with long lifetimes, so that we may see the
highly-ionizing charged tracks 
at the hadron or $e^+ e^-$ linear colliders \cite{Gherghetta:1999sw,
Chen:1995yu}.
We discuss a scenario with the wino dark matter in the next subsection.


The little fine-tuning problem in the MSSM is left in the
model~\cite{Casas:2003jx, Harnik:2003rs}. The non-observation of the Higgs boson
requires relatively heavy stops, indicating a high SUSY breaking
scale. A certain degree of fine-tuning is necessary to obtain the
correct size of the Higgs VEV.
In the NMSSM, the $\lambda S H_1 H_2$ coupling may raise the Higgs boson
mass without upsetting the perturbativity if $\lambda \lesssim 0.7$.
However, in the model with anomaly mediation, it is of no help since the
$\lambda$ parameter is necessary to be as small as $0.3$.

\subsection{Wino cold dark matter}


%
The wino is a perfect candidate for the dark matter of the universe
despite its large annihilation cross section \cite{Gherghetta:1999sw,
Moroi:1999zb}.
The current abundance can be explained by the non-thermal production
from the decay of the gravitinos which were produced when the universe
had high temperatures.
In the following, we discuss the relic mass density of the wino
including the effects of the non-thermal production.

By assuming the instantaneous decay of the gravitino at the decay
temperature $T_d$, the yield\footnote{The yield is defined relative to
  the entropy density.}  of the wino from the gravitino decay is equal
to that of the gravitino at $T_d$~\cite{Kawasaki:1994af,Bolz:2000fu}
and is given by
\begin{eqnarray}
 Y_{\chi}^{\rm NT}(T_d) \simeq Y_{3/2}(T_{RH}) \simeq
1.9 \times 10^{-12} \times\left(\frac{T_{RH}}{10^{10}{\rm GeV}}\right),
\label{eq:winont}
\end{eqnarray}
where $T_{RH}$ is the reheating temperature of the universe.
The decay temperature $T_d$ is given by
\begin{eqnarray}
 T_d \simeq 9\,{\rm MeV} \left(\frac{10}{g_{*}(T_d)}\right)^{1/4}
\left(\frac{m_{3/2}}{100{\rm TeV}}\right)^{3/2} 
\simeq 0.8\,{\rm MeV}\left(\frac{10}{g_{*}(T_d)}\right)^{1/4}
\left(\frac{m_{\rm wino}}{100{\rm GeV}}\right)^{3/2}, 
\end{eqnarray}
where $g_{*}(T)$ denotes the number of the effective massless degrees of
freedom and we use $m_{\rm wino} \simeq 5.2 \times 10^{-3} m_{3/2}$ (see
Eq.~(\ref{eq:soft})).
In the limit of the pure wino LSP, the dominant annihilation process is
the pair annihilation into two $W$-bosons via $t$-channel charged wino
exchange diagram.
Because $T_d$ is much lower than the wino mass, we can take the
non-relativistic limit of the cross section as a good approximation and
that is given by \cite{Moroi:1999zb} 
%
%
%
\begin{eqnarray}
\left\langle{\sigma v}\right\rangle = 
\frac{g_2^4}{2\pi}\frac{1}{m_{\rm wino}^2} 
\frac{(1-m_W^2/m_{\rm wino}^2)^{3/2}}{(2-m_W^2/m_{\rm wino}^2)^2}.
\end{eqnarray}
With the annihilation cross section, we obtain the yield $Y_\chi$ at low
temperature $T$ by solving the Boltzmann equation as
follows~\cite{Fujii_Hamaguchi1}:
%
\begin{equation}
Y_{\chi}(T)=
\left[\frac{1}{Y_{\chi}^{\rm TH}(T_{d})+Y_{\chi}^{\rm NT}(T_{d})}
+\frac{1}{Y_{\chi}^{\rm ann}(T_{d},T)}\right]^{-1}.
\label{eq:winoyield}
\end{equation}
Here, $Y_{\chi}^{\rm TH}(T_{d})$ is the thermal relic of the wino, and
$Y_{\chi}^{\rm ann}(T_{d},T)$ represents the annihilation effects after
the non-thermal production.
Those quantities are given by
\begin{eqnarray}
 Y_{\chi}^{\rm TH}(T_d) &\simeq &
10^{-14}\times\left(\frac{m_{\rm wino}}{100{\rm GeV}}\right),\\
\label{eq:winoth}
Y_{\chi}^{\rm ann}(T_d,T) 
& \simeq &
\sqrt{
\frac{45}{8 \pi^2 g_* (T_d)}
}
\frac{1}{\langle \sigma v \rangle M_{\rm Pl} (T_d - T)} \\
&\simeq&  2 \times 10^{-10}\times 
\left(\frac{g_{*}(T_{d})}{10}\right)^{-1/4}
\left(\frac{m_{\rm wino}}{100{\rm GeV}}\right)^{1/2}.
\label{eq:winoann}
\end{eqnarray}
We can neglect the second term in Eq.~(\ref{eq:winoyield}) for $T_{RH}
\lesssim 10^{12}$~GeV, since $Y^{\rm NT}_\chi (T_d)$ is much smaller than
$Y^{\rm ann}_\chi (T_d, T)$.
In this range, the yield $Y_\chi$ and the mass density parameter of the
wino $\Omega_\chi$ are simply given by
%
%
 \begin{eqnarray}
 Y_{\chi}(T) \simeq 
10^{-14}\times\left(\frac{m_{\rm wino}}{100{\rm GeV}}\right) 
+ 1.9 \times 10^{-12} \times\left(\frac{T_{RH}}{10^{10}{\rm GeV}}\right),
\label{eq:winofinal}
\end{eqnarray}
 \begin{eqnarray}
\nonumber
 \Omega_{\chi} h^2 
&=&\frac{m_{\rm wino}Y_{\chi}}{3.6221 \times 10^{-9}\,{\rm GeV}} \\
&\simeq& 
2.8\times 10^{-4}\times\left(\frac{m_{\rm wino}}{100{\rm GeV}}\right)^2 
+ 5.3 \times 10^{-2} \times\left(\frac{m_{\rm wino}}{100{\rm GeV}}\right)
\left(\frac{T_{RH}}{10^{10}{\rm GeV}}\right).
\label{eq:density}
\end{eqnarray}
Eq.(\ref{eq:density}) indicates that the wino can be the dominant
component of the dark matter (i.e. $\Omega_{\chi}h^2\simeq 0.11$) for an
appropriate reheating temperature.

\begin{figure}[t]
\begin{center}
\includegraphics[height=5.5cm]{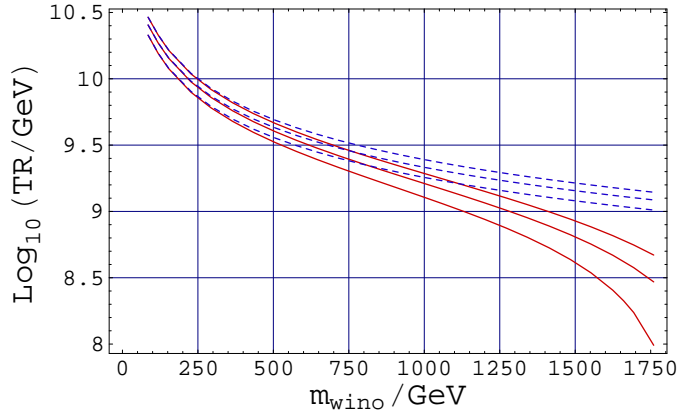} 
\caption{The required reheating temperature in order for the wino to be
the dark matter.  The solid lines corresponds to the required reheating
temperature for $\Omega_{\chi}h^2\simeq 0.095, 0.11, 0.13$ and the
dashed lines for $\Omega_{\chi}^{\rm NT}h^2\simeq 0.095, 0.11, 0.13$
from the bottom to the top, respectively.  } \label{fig:winoDM}
\end{center}
\end{figure}       

We show in Fig.~\ref{fig:winoDM} the required reheating temperature of
the universe as a function of the wino mass in order for the wino to be
the dark matter.
Solid lines correspond to $\Omega_{\chi}h^2\simeq 0.095, 0.11, 0.13$
from the bottom to the top, respectively.
We also plot in the same figure the reheating temperature which satisfy
$\Omega_{\chi}^{\rm NT}h^2= m_{\rm wino}Y_\chi^{\rm NT}/(3.6221\times
10^{-9}\,{\rm GeV})\simeq 0.095, 0.11, 0.13$ as dashed lines. The mass
density $\Omega_{\chi}^{\rm NT}h^2$ is the component of the non-thermally
produced wino through the gravitino decay.
%
%
As we see, $\Omega_\chi^{\rm NT}$ dominates the mass density of the dark
matter for $m_{\rm wino}\lesssim 500$~GeV.

Remarkably, the required reheating temperature is consistent with the
lower bound on $T_{RH}$ for the thermal leptogenesis, $T_{RH} \gtrsim 4
\times 10^9$~GeV \cite{Giudice:2003jh,Buchmuller:2004nz}. The lower
bound indicates the upper bound on the wino mass to be $m_{\tilde{W}}
\lesssim 500$~GeV.
%

\subsection{A solution to the strong CP problem}
\label{sec:strongCP} 

The UV insensitive anomaly mediation has a potential to solve the strong
CP problem by the Nelson-Barr \cite{Nelson:1983zb,Barr:qx}.
The mechanism is claimed not to be a good solution in SUSY models
because the SUSY breaking effect reintroduces the strong CP phase
$\bar{\theta}$ at one-loop level~\cite{Dine:1993qm}. However, with the
UV insensitivity of the SUSY breaking terms, it is revived as a solution to
the problem (see also \cite{Hiller:2001qg,Hiller:2002um}).

The mechanism needs CP to be an exact symmetry of the Lagrangian such
that the $\bar{\theta}$ parameter as well as the phase in the CKM matrix
vanish.
The observed non-vanishing CP phase in the CKM matrix can be induced at
low energies by the spontaneous CP violation. Nelson and Barr proposed
models in which we obtain only the CKM phase by the spontaneous CP
violation while $\bar{\theta}$ remains vanishing at tree level.
The simplest model is the following. We introduce a pair of vector-like
quarks $d^c_4$ and $\bar{d^c_4}$ which have the same and opposite
quantum numbers as the right-handed down-type quarks. With the
superpotential
\begin{eqnarray}
 W = a X d^c_i \bar{d}^c_4  + f_d^{ij} q_i H_d d_j^c + m d^c_4 \bar{d}^c_4\ , 
\end{eqnarray}
where $i,j=1,2,3$ and $\alpha=1-4$, the phase in the VEV of $X$ induce
the CKM phase, but the determinant of the mass matrix remains to be real
indicating no contribution to $\bar{\theta}$.

In general, such new matter fields with flavor-dependent couplings
reintroduce the SUSY flavor problem through their loops.  It is the
non-trivial virtue of the UV insensitivity that decouples these
interactions from the low-energy soft terms so that we can discuss
such additional interactions without conflicting the phenomenological
limits.  In turn, the UV insensitivity turns out to be crucial for the
mechanism as well as we discuss below.

Once we realize the situation with the non-zero CKM phase and vanishing
$\bar{\theta}$ at tree level, the non-renormalization theorem in SUSY
ensures vanishing $\bar{\theta}$ even at quantum level.
Potentially problematic is the SUSY breaking effect as mentioned before.
One-loop diagrams with SUSY breaking couplings induce the $\bar{\theta}$
parameter and acceptably small values of $\bar{\theta}$ require
extraordinarily high degrees of degeneracy among SUSY breaking
parameters~\cite{Dine:1993qm}. 
However, the UV insensitive anomaly mediation as well as gauge
mediation~\cite{Barr:1996wx} does not suffer from the problem.
The UV insensitivity ensures that all the SUSY breaking terms are
described by the SUSY invariant quantities such as the Yukawa and gauge
coupling constants, and therefore there is no new Jarlskog invariant
other than that of the SM, i.e., $\Im(\det [m_u m_u^\dagger, m_d
m_d^\dagger])$.
With the small Jarlskog invariant and significant loop suppression
factors, the $\bar{\theta}$ parameter remains to be very small such as
$10^{-29} - 10^{-19}$~\cite{Hiller:2002um}, which is much
smaller than the experimental upper bound of $10^{-10}$.

\subsection{The cosmological domain wall and the tadpole problems}
%
One may worry about the formation of the cosmological domain wall
associated with the $Z_3$ symmetry breaking in the
NMSSM~\cite{Abel:1995wk}.
However, with the presence of the vector-like quarks $D$ and $\bar{D}$,
there is no domain wall problem.  Since the $Z_3$ symmetry, under which
all the chiral superfields has a unit charge, is anomalous with respect
to the SU(3)$_C$ gauge group, the instanton effect can give a sufficient
energy shift among domains such that the domain wall is
unstable~\cite{Preskill:1991kd}.

Related to this issue, there is a problem of the instability of the
Higgs potential caused by a loop correction associated with the SUSY
breaking effect. The tadpole diagrams of $S$ with gravitational
interactions diverge quadratically, and reintroduce the hierarchy
problem~\cite{Abel:1995wk, Bagger:1995ay}.
The appearance of the linear term can be forbidden if the gravitational
interaction respects the $Z_3$ symmetry. It is, however, argued that the
quantum gravity violates all the global symmetry. Since the $Z_3$
symmetry is anomalous, we cannot think of the symmetry as a gauge
symmetry, and thus it is expected to be broken at the Planck scale.

Although it may not be a problem once we understand quantum gravity, it
is possible to control the tadpole divergence by embedding $Z_3$ to a
higher anomaly free symmetry $Z_{3N}$ such that the $Z_3$ symmetry is
realized as an accidental approximate symmetry after breaking of
$Z_{3N}$.
For example, the $Z_{3N}$ symmetry, under which all the chiral
superfields have charge $N$, can be made anomaly free when we add $N$
pairs of vector-like quarks ${\cal D}$ and ${\cal \bar{D}}$ which have
the same quantum numbers as $D$ and $\bar{D}$ under the SM gauge group
and charge $-1$ under $Z_{3N}$.
When the $Z_{3N}$ symmetry is broken spontaneously by the VEV of a field
$\Sigma$ which has $Z_{3N}$ charge 2, the vector-like fields acquire
the mass of the order of $\langle \Sigma \rangle$ and decouple from the
low-energy physics if the VEV is much larger than the electroweak scale.
The mass term of the vector-like quarks $D$ and $\bar{D}$ or the
$\mu$-term given through the VEV of $\Sigma$ is naturally small by the
$Z_{3N}$ symmetry, because it restricts the form of the lowest
dimensional interaction to be $\Sigma^X / M_{\rm Pl}^{X-1} D \bar{D}$
where $X=N/2$ or $2N$ for even or odd $N$, respectively.
The tadpole term in the Lagrangian is also suppressed as $m_{3/2}
\langle \Sigma \rangle^N / M_{\rm Pl}^{N-2} S$, which can be small
enough.
Also, the $Z_{3N}$ symmetry may naturally forbids the large mixing
between ordinary quarks $d^c$ and $\bar{D}$. For example, if the
$Z_{3N}$ charges of $D$ and $\bar{D}$ are $N+n$ and $N-n$, respectively,
the mixing is naturally suppressed by the $Z_{3N}$ breaking effect. The
small but finite mixing makes the vector-like fields unstable, which
ensures the absence of the problem with overclosure of the universe
\cite{Han:1997wn}.
%

\section{Conclusions}
\label{sec:conclusion}

We have examined electroweak symmetry breaking in the NMSSM with UV
insensitive anomaly mediation. We added a vector-like pair of matter
fields to solve the light Higgsino problem in the NMSSM with the anomaly
mediation.
Viable parameter regions are found to preserve the perturbative
coupling unification.

With the success of the electroweak symmetry breaking, the model is a
phenomenologically and cosmologically perfect package for particle
physics.
The SUSY FCNC and CP problems are solved thanks to the UV insensitivity
of the anomaly mediation. Also, the strong CP phase is not induced by
the SUSY breaking effect once we set vanishing $\bar{\theta}$ at tree
level, which is claimed to be natural in the string theory context
\cite{Dine:1992ya}.
The lightest neutralino is the wino which is a good candidate of the
cold dark matter assuming the non-thermal production from the gravitino
decay.
The thermal leptogenesis works without contradicting the constraint
on the gravitino abundance from the BBN theory, since the
gravitino decays before the BBN era.
There is no cosmological domain wall problem associated with the
spontaneous $Z_3$ symmetry breaking in the NMSSM because the $Z_3$
symmetry is broken by the QCD instanton effect in the presence of the
vector-like quarks.
The tadpole problem in the NMSSM may be avoided by imposing anomaly
free $Z_{3N}$ symmetry at high energy.

\section*{Acknowledgments}

We thank Tsutomu Yanagida for useful discussions on the wino dark matter
and the constraint on the reheating temperature of the universe in the
scenario, and Raman Sundrum for encouragements.
We also thank Zackaria Chacko for discussions and comments on the
early draft of the paper.  HM thanks the Institute for Advanced Study
where the work was initiated.
This work was supported by the Institute for Advanced Study, funds for
Natural Sciences, as well as in part by the DOE under contracts
DE-AC03-76SF00098 and in part by NSF grant PHY-0098840.

\appendix
\section{Anomalous dimensions and beta functions}
\label{app:rge}

We list the anomalous dimensions and the beta functions in the model
which are necessary to compute the soft SUSY breaking terms.

The anomalous dimensions are given by
\begin{eqnarray}
 (4 \pi)^2 \gamma_{H_1} =
f_\tau^2 + 3 f_b^2 + \lambda^2 - \frac{1}{2} g_Y^2 - \frac{3}{2} g_2^2\ ,
\end{eqnarray}
\begin{eqnarray}
 (4 \pi)^2 \gamma_{H_2} =
3 f_t^2 + \lambda^2 - \frac{1}{2} g_Y^2 - \frac{3}{2} g_2^2 \ ,
\end{eqnarray}
\begin{eqnarray}
 (4 \pi)^2 \gamma_{S} =
2 \lambda^2 + 2 h^2 + 3 k_D^2 + 2 k_L^2\ ,
\end{eqnarray}
\begin{eqnarray}
 (4 \pi)^2 \gamma_{l_i} =
f_\tau^2 \delta_{i3} - \frac{1}{2} g_Y^2 - \frac{3}{2} g_2^2  \ ,
\end{eqnarray}
\begin{eqnarray}
 (4 \pi)^2 \gamma_{e^c_i} =
2 f_\tau^2 \delta_{i3} - 2 g_Y^2  \ ,
\end{eqnarray}
\begin{eqnarray}
 (4 \pi)^2 \gamma_{q_i} =
( f_b^2 + f_t^2 ) \delta_{i3} 
- \frac{1}{18} g_Y^2 - \frac{3}{2} g_2^2  - \frac{8}{3} g_3^2  \ ,
\end{eqnarray}
\begin{eqnarray}
 (4 \pi)^2 \gamma_{d^c_i} =
2 f_b^2  \delta_{i3} 
- \frac{2}{9} g_Y^2  - \frac{8}{3} g_3^2  \ ,
\end{eqnarray}
\begin{eqnarray}
 (4 \pi)^2 \gamma_{u^c_i} =
2 f_t^2  \delta_{i3} 
- \frac{8}{9} g_Y^2  - \frac{8}{3} g_3^2  \ ,
\end{eqnarray}
\begin{eqnarray}
 (4 \pi)^2 \gamma_{D} =
k_D^2
- \frac{2}{9} g_Y^2  - \frac{8}{3} g_3^2  \ ,
\end{eqnarray}
\begin{eqnarray}
 (4 \pi)^2 \gamma_{\bar{D}} =
k_D^2
- \frac{2}{9} g_Y^2  - \frac{8}{3} g_3^2  \ ,
\end{eqnarray}
\begin{eqnarray}
 (4 \pi)^2 \gamma_{L} =
k_L^2
- \frac{1}{2} g_Y^2  - \frac{3}{2} g_2^2  \ ,
\end{eqnarray}
\begin{eqnarray}
 (4 \pi)^2 \gamma_{\bar{L}} =
k_L^2
- \frac{1}{2} g_Y^2  - \frac{3}{2} g_2^2  \ ,
\end{eqnarray}
where we neglect the Yukawa coupling constants for the first and second
generations.

The beta functions for the gauge coupling constants are the following:
\begin{eqnarray}
 ( 4 \pi )^2 \frac{d g_Y}{d \ln \mu} = \frac{38}{3} g_Y^3\ ,
\end{eqnarray}
\begin{eqnarray}
 ( 4 \pi )^2 \frac{d g_2}{d \ln \mu} = 2 g_2^3\ ,
\end{eqnarray}
\begin{eqnarray}
 ( 4 \pi )^2 \frac{d g_3}{d \ln \mu} = - 2 g_3^3\ .
\end{eqnarray}

The beta functions for the Yukawa coupling constants are expressed by
the anomalous dimensions as follows:
\begin{eqnarray}
 ( 4 \pi )^2 \frac{d f_\tau}{d \ln \mu} & = &  
 ( 4 \pi )^2 f_\tau ( \gamma_{l_3} + \gamma_{H_1} + \gamma_{e^c_3})
\notag \\
 & = &  f_\tau 
( 4 f_\tau^2 + 3 f_b^2 + \lambda^2 - 3 g_Y^2 - 3 g_2^2 )\ ,
\end{eqnarray}
\begin{eqnarray}
 ( 4 \pi )^2 \frac{d f_b}{d \ln \mu} &=&
 ( 4 \pi )^2  f_b ( \gamma_{q_3} + \gamma_{H_1} + \gamma_{d^c_3}) 
\nonumber \\
& = & f_b
( f_\tau^2 + 6 f_b^2 + f_t^2 + \lambda^2
-\frac{7}{9} g_Y^2 - 3 g_2^2 - \frac{16}{3} g_3^2 )
\ ,
\end{eqnarray}
\begin{eqnarray}
 ( 4 \pi )^2 \frac{d f_t}{d \ln \mu} & = &
 ( 4 \pi )^2  f_t ( \gamma_{q_3} + \gamma_{H_2} + \gamma_{u^c_3})
\nonumber \\
& = & f_t
( 6 f_t^2 + f_b^2 + \lambda^2 
-\frac{13}{9} g_Y^2 - 3 g_2^2 - \frac{16}{3} g_3^2 )
\ ,
\end{eqnarray}
\begin{eqnarray}
  ( 4 \pi )^2 \frac{d \lambda}{d \ln \mu} &=&
  ( 4 \pi )^2  \lambda ( \gamma_S + \gamma_{H_1} + \gamma_{H_2})
\nonumber \\
& = &
\lambda 
( f_\tau^2 + 3 f_b^2 + 3 f_t^2 + 4 \lambda^2 + 2 h^2 + 3 k_D^2 + 2 k_L^2 
- g_Y^2 - 3 g_2^2 )
\ ,
\end{eqnarray}
\begin{eqnarray}
 ( 4 \pi )^2 \frac{d h}{d \ln \mu} &=&
 ( 4 \pi )^2  3 h \gamma_S 
\nonumber \\
&=&
3 h ( 2 \lambda^2 + 2 h^2 + 3 k_D^2 + 2 k_L^2  )
\ ,
\end{eqnarray}
\begin{eqnarray}
 ( 4 \pi )^2 \frac{d k_D}{d \ln \mu} &=&
 ( 4 \pi )^2  k_D ( \gamma_S + \gamma_D + \gamma_{\bar{D}})
\nonumber \\
&=&
k_D ( 2 \lambda^2 + 2 h^2 + 5 k_D^2 + 2 k_L^2 
- \frac{4}{9} g_Y^2 - \frac{16}{3} g_3^2 )
\ ,
\end{eqnarray}
\begin{eqnarray}
 ( 4 \pi )^2 \frac{d k_L}{d \ln \mu} &=&
 ( 4 \pi )^2  k_L ( \gamma_S + \gamma_L + \gamma_{\bar{L}})
\nonumber \\
&=&
k_L ( 2 \lambda^2 + 2 h^2 + 3 k_D^2 + 4 k_L^2 
- g_Y^2 - 3 g_2^2)
\ .
\end{eqnarray}


\end{document}